\documentclass[preprint2]{aastex}

\shorttitle{Kick-off black hole?} \shortauthors{Dottori
et al.}

\begin{document}

\title{Is J\,133658.3--295105 radiosource at z\,$\geq$\,1 or at the distance of M83?}

\author{Horacio Dottori\altaffilmark{1}, Rub\'en J. D\'{\i}az\altaffilmark{2,3}, Dami\'an Mast\altaffilmark{3,4}}

\altaffiltext{1}{Instituto de F\'{\i}sica -- Universidade Federal do
Rio Grande do Sul, Porto Alegre, Brazil. e-mail:dottori@if.ufrgs.br}

\altaffiltext{2}{Gemini Observatory, Southern Operations
Center, La Serena, Chile.}

\altaffiltext{3} {Consejo Nacional de Investigaciones
Cient\'{\i}ficas y Tecnol\'ogicas, Argentina.}

\altaffiltext{4}{Observatorio Astron\'omico, Universidad
Nacional de C\'ordoba, Argentina.}

\begin{abstract}
 We present Gemini optical imaging and spectroscopy of the
radio source J\,133658.3--295105. This source has been
suggested to be the core of an FRII radio source with two
detected lobes. J\,133658.3--295105 and its lobes are
aligned with the optical nucleus of M\,83  and with  three
other radio sources at the M\,83 bulge outer region. These
radio sources are neither supernova remnants nor HII
regions. This curious configuration prompted us to try to
determine the distance to J\,133658.3--295105.  We detected
H$\alpha$ emission redshifted by
$\approx$\,130\,km\,s$^{-1}$ with respect to an M\,83 HII
region 2\farcs5\, E-SE of the radio source. We do not
detect other redshifted emission lines of an optical
counterpart down to $m_i=22.2\pm0.8$. Two different
scenarios are proposed: the radio source is at
$z\geqslant2.5$, a much larger distance than the previously
proposed lower limit $z\geqslant1.0$, or the object was
ejected by a gravitational recoil event from the M\,83
nucleus.  This nucleus is undergoing a strong dynamical
evolution judging from previous 3D spectroscopy.

\end{abstract}



\section{INTRODUCTION}

The radio source J\,133658.3--295105 (object 28 in
\cite{maddox06} list) in M\,83, is described as the core of
an FR II radio source \cite[]{maddox06, soria03} whose
radio lobes are objects 27 and 29 in the same list. The
source presents characteristics of a radio galaxy or a
quasar at a distance $z\geqslant 1$ as discussed by
\cite{soria03}. These authors have found that at a distance
d$_L$\,=\,4.7\,Gpc ($z$\,=1, q$_0$\,=\,0.5,
H\,=\,75\,km\,s$^{-1}$\,Mpc$^{-1}$), J\,133658.3--295105
$L_{20\,cm}$ and $L_{6\,cm}$  fit reasonably well with the
expected FR\,II radio-sources luminosities.

\begin{figure}
\centerline{ \scalebox{0.38}{
\includegraphics{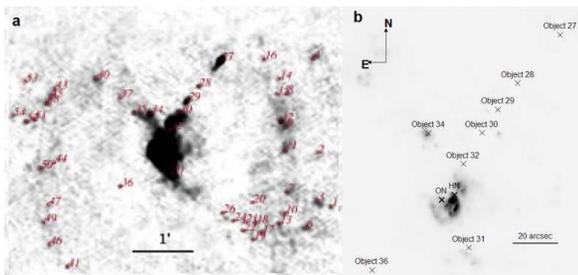}
} }
  \caption{Left: Radiomap of M\,83 6\,cm emission, from
  \cite{maddox06}.
  Right: dettached line of radiosources, including the galaxy
optical nucleus.}\label{fig:global}
\end{figure}

M\,83 and NGC\,5128 (Cen\,A) are the largest galaxies in
one of the most active regions in the nearby sky, the
Hydra-Centaurus group. Merger processes as well as Galactic
inner build-up are evident in both galaxies. The radio
source (RS) 28 appears projected at about 60\arcsec\, from
the nucleus of M\,83 and the line joining RSs 27, 28, and
29 appears to emerge from the M\,83 bulge as seen in
Figure~\ref{fig:global} \cite[and Figure\,1
of][]{maddox06}. This alignment was originally proposed by
\cite{cowan94} to be a jet from the nucleus of M\,83.
Moreover, RSs 30, 32, and 36, and the M\,83 optical nucleus
(ON) itself appears to be strongly aligned with this
structure, too (Figure~\ref{fig:global}, right) The whole
structure covers $\sim100$\arcsec\ on the sky. None of
these radio sources are supernova remnants nor do they
appear associated with HII regions \citep{maddox06}. RS 32
emits X-rays as a soft X-ray binary. RS 36 was also
described as an X-ray binary with evidence for X-ray
spectral evolution. The optical nucleus of M\,83 is also a
radio and X-ray source \citep{soria03}.

These intriguing features encouraged us to try to determine
more precisely the distance to RS 28. The complex structure
of the central 300\,pc of the bulge of M\,83 suggests the
possibility of it hosting a kick-off object, which has been
theoretically predicted \citep{libeskind06, gualandris07,
bonning07}, but not yet observed \citep{dottori07}.

Indeed, the condensation traditionally recognized as the
M\,83 optical nucleus (ON) (Figure~\ref{fig:kick-off}c) is
off-centered by $4\arcsec$\,($\sim$\,80\,pc) to the
northeast with respect to the center of the external
$K$-band isophotes of the bulge \citep{jen81}. At the bulge
center, \cite{thatte00} have found the kinematic center
(KC). More recently, \cite{mast06} demonstrated the
existence of a third condensation (hidden nucleus,HN), also
hidden at optical wavelengths,
7$\arcsec$\,($\sim$\,140\,pc) to the west-southwest of KC
which seems to be a small galaxy being cannibalized
\citep{diaz06b}. There is an arclet $\approx\,7\arcsec$ to
the southwest of KC (Figure~\ref{fig:kick-off}c) with more
than twenty 30-Dor-like giant HII regions \citep{Harris01}.
Pa$\beta$ observed around KC and HN indicates the presence
of regions that are actively forming stars \citep{diaz06a}.

The derived mass of ON is $\sim4\times\,10^6\,M_{\odot}$
\citep{debra97, thatte00}. These authors proposed a similar
mass for KC, while \cite{dottori07} determined a larger
value of $\approx$\,60 $\times10^6$\,M$_{\odot}$. HN has a
mass of 16$\times10^6$\,M$_{\odot}$ \citep{diaz06a}. The
upper mass limits for the putative black holes (BHs)
associated with ON, KC, and HN are 0.2--1.0\,$\times10^6\,$
M$_\odot$ \citep{dottori07}.

 Numerical simulations  show
that ON, KC, HN and the arclet of HII regions will merge,
forming a single massive core in a few hundred Myr
\citep{diaz06b}. Parallel to the merging process at the
center of M 83, molecular clouds in the bar seem to be
funneled inward \citep{debra97}, guaranteeing central mass
accretion in the future.

We discuss here the optical observations of the radio
source J\,133658.3--295105, taken to more precisely derive
its distance and its nature.

\section{OBSERVATIONS}

\begin{figure}
\centerline{ \scalebox{0.4}{
\includegraphics{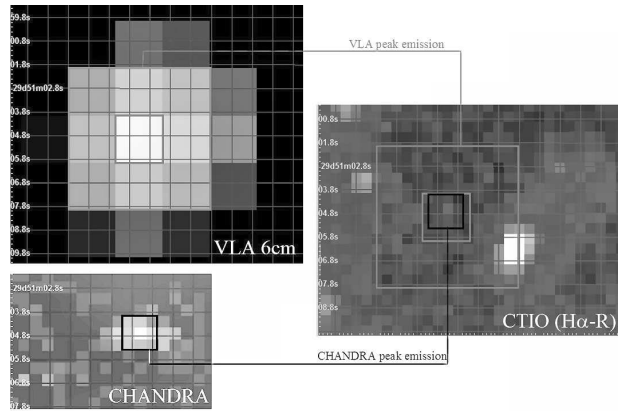}
} }
  \caption{Astrometry around RS\,28  showing the relative
spatial resolution of CTIO (H$\alpha$-R), 6\,cm radio map
and Chandra's X-ray map. The brightest knot in the CTIO
(H$\alpha$-R) image is the HII region at 2\farcs5\,
east-southeast of RS}\label{fig:RS28_position}
\end{figure}

In 2007 April, we obtained a 5 hr spectrum with the REOSC
echelle spectrograph in a simple dispersion mode at the
CASLEO  2.2\,m telescope, in Argentina. REOSC has a
1024$\,\times\,$1024 TEK CCD, with a 24$\mu$m pixel size. A
1200 lines mm$^{-1}$ grating was used, covering the
wavelength range from 6200 to 6900 \AA\, around
H$_{\alpha}$. The dispersion was 32 \AA\,mm$^{-1}$, the
reciprocal dispersion 0.76 \AA\,pix$^{-1}$, the resolution
2.5 \AA, and the angular scale was 1\farcs02 pix$^{-1}$.
The slit (at P.A.\,=\,152$^\circ$) was positioned on an HII
region 2\farcs5 to the to the east-southeast of RS\,28.
These 2\arcsec\ seeing observations confirmed the nature of
the HII region. On the basis of an accurate astrometric
analysis of the source in different spectral domains, we
planned the subsequent Gemini Multi Object
Spectrographs-South (GMOS-S) observations.
Figure~\ref{fig:RS28_position} shows the astrometry and the
HII region to the east-southeast of RS\,28 on a
(H$\alpha\,\--\,R$) Cerro Tololo Inter-American Observatory
(CTIO) image.

RS 28 was observed by GMOS-S at GEMINI South in 2007 June
(program GS-2007A-DD-17). We performed broad and narrow
band imaging in the range 500 to 950\,nm, centered at the
position of the radio source emission peak with average
seeing of $0\farcs8$. The observed bands and exposure times
were $g$: 90\,s; [OIII]: 121\,s; [OIII]C: 121\,s;
H$\alpha$: 91.5\,s; H$\alpha$C: 91.5\,s; [SII]: 91.5\,s;
$i$: 60\,s; CaT: 60\,s and $z$: 91.5\,s. Spectroscopy in
the spectral range  572-988\,nm\, was performed  with an
effective exposure time of 3600\,s and average seeing of
$0\farcs8$. A $1\farcs5$ long slit was positioned on RS\,28
and oriented along the line of radio sources shown in
Figure~\ref{fig:global} (P.A.\,=\,140$^\circ$).

\section{WHERE IS RS 28?}

\begin{figure}
\centerline{ \scalebox{0.4}{
\includegraphics{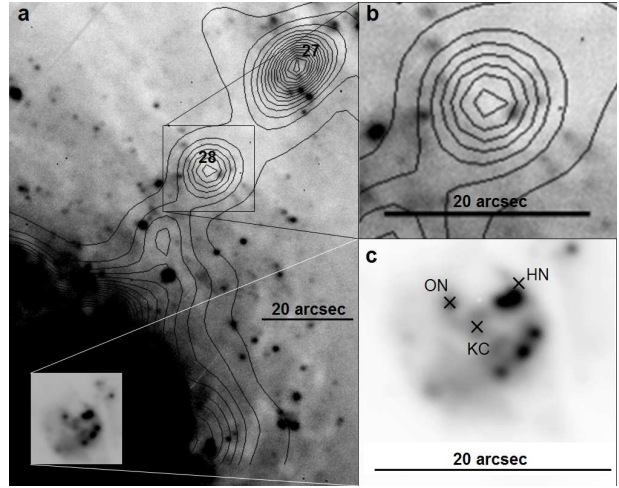}
} }
  \caption{a) GMOS  500\,nm to 950\,nm co-added image superimposed on the
radioisophotes around RS\,28.  b) Image blow-up around RS\,28
showing that no optical counterpart is detected.  c) Blow up of the
bulge center in the co-added image showing the region complexity.
The position of the optical nucleus (ON), kinematic center (KC) and
hidden nucleus (HN) are denoted with crosses.}\label{fig:kick-off}
\end{figure}

Figure~\ref{fig:kick-off} shows the co-added image from the
observed GMOS-S bands $V$, [OIII], [OIII]C, H$\alpha$,
H$\alpha$C, [SII], $z$, $i$, and CaT. We estimate that the
detection limit at signal-to-noise ratio ($S/N)\sim3$ is
m$_i=23.5\pm0.5$. This is based on an average seeing of
$0\farcs8$, and takes into account the background noise
from the M\,83 disk.

It is expected that the disk of M\,83 will dim background
objects. \cite{boissier05} derive an azimuthal mean
absorption $A(FUV)=2.0$\,mag at R around 1\,kpc, which
translates into $A(I)=1.3$\,mg (assuming
$A(FUV)/A(I)=1.5$). \cite{beckman96} obtained, for three
face-on galaxies, an inter-arm absorption 0.2--0.5\,mag
smaller than the azimuthal mean at 800\,nm. That would lead
to an absorption of the order of $A(I)=1.0\pm0.3$\,mag for
background objects at the projected position of RS\,28 on
the M\,83 disk. These considerations lead us to estimate
that our composed image could detect more than 80\% of the
broad- and narrow-line Active Galactic Nuclei (AGNs) at
$1\leq z \leq 3$ in the SEXSI sample \cite[see Figure\,6
in][]{eckart06}, irrespective of the redshift. The blow up
of the composed GMOS image of the 20\arcsec\, around RS 28
in Figure~\ref{fig:kick-off}\,b does not show any sign of
an associated optical source.

We have determined the detectability of RS 28 in our images
using Cen A as an example. The radio source Cen A
originates from the elliptical galaxy NGC\,5128, which
hosts a Seyfert\,2 nucleus with a strong dust lane across
its main body. Quasars with strong activity of star
formation would present much dust, and probably would
suffer internal absorption as NGC\,5128 does. The absolute
brightness of NGC\,5128 is M$_B$\,=\, -23.0\,mag at a
luminosity distance 11\,Mpc, $(m-M)\,=\,30.2$.  If Cen A
was located at $z\sim1$, NGC\,5128 would shine at
m$_I$\,=\,21.5\,mag, or m$_I$\,=\,23.1\,mag if located at
$z=2$. Therefore, if located somewhere in the redshift
range $1\leq z\leq2$, a Cen A type galaxy would be
detectable by our imagery. The angular distance between the
maxima of Cen A radio lobes brightness is $\sim$\,3$^\circ$
(580\,kpc) about  twice the value quoted by \cite{soria03}
for RS\,28 at $z=1$. That relation is maintained if RS\,28
is at larger $z$, since the angular size distance does not
change significantly with $z$, for $z\geq1$.

Our spectroscopy suggests that we should detect any distant
background AGN-typical emission lines. In the co-added
spectrum, the detection limit for the continuum of a giant
elliptical galaxy at redshift 2 is m$_i=20.5\pm0.8$
(S/N$\sim3$), and it is m$_i=22.2\pm0.8$ for H$\gamma$
detection in a QSO at $1<z<2$. Strong redshifted lines
should fall into the spectral range covered by GMOS for
$1\leq z \leq2.5$, as can be seen in the revised catalog of
QSOs \citep{hewitt93}. The most conspicuous emission lines
and the corresponding $z$ intervals are: [OII]
$\lambda$\,3727\, (0.34,1.4); [NV] $\lambda$\,3426\,
(0.46,1.77); [MgII] $\lambda$\,2798\, (0.8,2.4); [CIII]
$\lambda$\,1909\, (1.6,3.9); [HeII] $\lambda$\,1640\,\AA\,
(2.1,4.7); [CIV] $\lambda$\,1549\, (2.2,5.1); [SIV]
$\lambda$\,1402\, (2.6,6.7); and [NeV] $\lambda$\,1240\,
(3.4,7.6). Our observational constrains set a limit of the
order of $z\approx 3$ for large radio galaxies.

We can get an idea of the expected S/N by looking at
spectra obtained for broad- and narrow-lines AGNs at
different distances by the Keck observatory with exposure
times of 1 to 2 hs \citep{eckart06}. For a similar exposure
time (3600\,s), we expect for GEMINI a S/N$\approx$0.7
times that of Keck. So, practically all lines, except for
[HeII] would be easily seen in our spectra. Therefore, from
our optical spectroscopy, we can confidently say that if
RS\,28 is a high-redshift radio source, it should be at
$z\geq 2.5$.

\begin{figure}[t]
\centerline{ \scalebox{0.5}{
\includegraphics{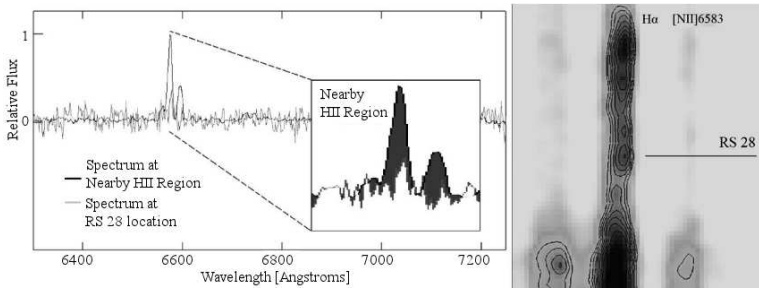}
} }
  \caption{Right: GEMINI+GMOS 2-D spectrum around RS\,28.
Left: $I$\,vs.\,$\lambda$ shows the redshifted H$\alpha$ at
RS\,28 with respect to that of the HII region located to
the N.}\label{fig:espectro1}
\end{figure}

The RS\,28 spectrum is compared in
Figure~\ref{fig:espectro1} with that of a simultaneously
observed, well-defined HII region located on the arm to the
north of RS\,28. We see that H$\alpha$ emission appears
marginally at the RS\,28 spectrum, redshifted with respect
to that of the HII region. Radial velocities are
respectively \,618$\,\pm$30\,km\,s$^{-1}$ and
522$\,\pm\,$12\,km\,s$^{-1}$. The HII region located to the
east-southeast of RS\,28 (Figure~\ref{fig:RS28_position})
is at V$_r$\,=\,490$\pm$\,12\,km\,s$^{-1}$, based on our
REOSC-CASLEO data. REOSC accuracy can be gaged by the
nuclear spectrum taken on the same night and instrument
configuration, which gives for the nucleus a
V$_r$\,=\,534$\,\pm$\,15\,km\,s$^{-1}$. The heliocentric
velocity of M\,83, from the NASA/IPAC Extragalactic
Database (NED), is V$_r$\,=\,513$\,\pm\,$5\,km\,s$^{-1}$.
Two-dimensional CO observations \citep{sakamoto04} provide
for the nuclear region a value of
V$_{LSR}=508\pm15$\,km\,s$^{-1}$ (error guess is from the
position accuracy in Figure\,3 of Sakamoto et al.). This
value has to be added to the LSR velocity of approximately
24\,km\,s$^{-1}$ (obtained with Ed Murphy's calculator in
http:$\setminus\setminus$fuse.pha.jhu.edu) to compare with
our measurement. We conclude that REOSC radial velocities
are reliable within the errors.

The radial velocity of RS\,28 is approximately
130$\,\pm\,$40\,km\,s$^{-1}$ larger than that of the close
east-southeast HII region. The nucleus and the two HII
regions show similar radial velocities because M\,83
presents a small inclination of 24$^\circ$ with respect to
the plane of the sky and the direction in discussion is at
about 130$^\circ$ from the galaxy major axis and
practically perpendicular to the bar
(Figure~\ref{fig:kick-off}) where noncircular motions are
detected \citep{sakamoto04}. This result leads us to
conclude that the H$\alpha$ emission marginally detected at
RS\,28 is emitted by an object that does not take part of
the disk kinematics, whose projected velocity in the radial
direction is of the order of 100\,km\,s$^{-1}$. These
arguments point to a possible physical connection between
RS\,28 and M\,83.

\section{FINAL REMARKS}

Chance alignments do occur. A striking case is that of the
two radio sources juxtaposed 1\arcmin\, to the northeast
and southwest of Scorpius (Sco) X-1, which for a long time
were believed to be radio lobes of Sco X-1, until
\cite{fomalont91} demonstrated by measuring Sco X-1 proper
motion that they are unrelated background sources. On the
one hand, deep optical imaging and spectroscopy on the
location of the radio source RS\,28 do not confirm the
high-redshift nature of this object, and on the other hand,
GMOS/GEMINI-S spectroscopy marginally shows the presence of
H$\alpha$ in emission at a redshift similar to that of the
galaxy but from a source kinematically well-differentiated
from disk objects.

If its local character is confirmed, the FRII radio source
RS\,28 would have a total projected size
$\approx$\,0.5\--1.0\,kpc. The possibility of RS\,28 being
ejected by a gravitational recoil, remnant of a previous
M\,83 nuclear merger, remains open. In that case, the
driving engine within RS\,28 that produced the ejecta
RS\,27 and RS\,29 would be an accretion disk like those
postulated for microquasars in the Galaxy and extragalactic
radio sources. The reason why RS\,27 and RS\,29 sit along
the apparent kick line remains unclear.

Physical properties of accreting BHs (temperature, ejecta
size, etc) scale with some power of the BH mass
\citep{mirabel99}. If ejecta size scales with the BH mass,
a direct comparison with the microquasar 1E1740.7-2942
(M$_{BH}\,\approx$ 10\,M$\odot$, size\,$\approx$\,5\,pc;
\citealp{mirabel92}) leads to a mass for a RS\,28 BH
$\sim$\,1000 to 2000\,M$\odot$. The characteristic
blackbody temperature for Eddington-limit accretion
\citep{rees84, mirabel99} should be
$T\,=\,10^7\,(M/M\odot)^{\--1/4}$ or about 1/3 to 1/5 of
that of a Galactic microquasar.

X-ray and/or near-IR spectroscopic observations will be
necessary to scan redshifts $2.5\leqslant z\leqslant5.0$,
and also around $z\sim0$, to unequivocally determine the
distance to and the nature of J\,133658.3--295105.

\section{Acknowledgements}

We acknowledge the support of the CONICET grant PIP 5697,
and H.D. acknowledges support from CNPq (Brazil). We thank
Jean-Rene Roy, Percy Gomez, Maria Paz Ag\"uero and Elena
Terlevich for discussions. We also thank Ms.~Pippa Clarke
for a careful reading of the manuscript. Finally we
acknowledge the Gemini Observatory for the allocation of
director's discretionary telescope time, through the
proposal GS-2007A-DD-17. The Gemini Observatory is operated
by the Association of Universities for Research in
Astronomy, Inc., under a cooperative agreement with the NSF
on behalf of the Gemini partnership: NSF (USA), STFC
(United Kingdom), NRC (Canada), ARC (Australia), MINCYT
(Argentina), CNPq (Brazil), and CONICYT (Chile). The
NASA/ESA Hubble Space Telescope is operated by AURA under
NASA contract NAS 5-26555.


\end{document}